\documentclass[twocolumn,aps,superscriptaddress]{revtex4}

\usepackage{dcolumn}
\usepackage{bm}
\usepackage{epsf}
\usepackage[dvipdfmx]{graphicx} 
\usepackage{color}
\usepackage {mathcomp}
\begin{document}

\title{Soft and hard x-ray orbital-resolved photoemission study of\par a strongly correlated Cd-Ce quasicrystal approximant}

%
\author{Goro Nozue}
\thanks{goro.nozue@uni-wuerzburg.de}
\thanks{Present address: Physikalisches Institut and W\"{u}rzburg-Dresden Cluster of Excellence ctd.qmat, Julius-Maximilians-Universit\"{a}t W\"{u}rzburg, 97074 W\"{u}rzburg, Germany}
\affiliation{Division of Materials Physics, Graduate School of Engineering Science, The University of Osaka, Toyonaka, Osaka 560-8531, Japan}
\affiliation{RIKEN SPring-8 Center, Sayo, Hyogo 679-5158, Japan}
\author{Hidenori Fujiwara} 
\affiliation{Division of Materials Physics, Graduate School of Engineering Science, The University of Osaka, Toyonaka, Osaka 560-8531, Japan}
\affiliation{RIKEN SPring-8 Center, Sayo, Hyogo 679-5158, Japan}
\affiliation{Spintronics Research Network Division, Institute for Open and Transdisciplinary Research Initiatives, The University of Osaka, Yamadaoka 2-1, Suita, Osaka, 565-0871, Japan}
\author{Satoru Hamamoto}
\affiliation{RIKEN SPring-8 Center, Sayo, Hyogo 679-5158, Japan}
\author{Miwa Tsutsumi}
\author{Akane Ose}
\affiliation{Division of Materials Physics, Graduate School of Engineering Science, The University of Osaka, Toyonaka, Osaka 560-8531, Japan}
\affiliation{RIKEN SPring-8 Center, Sayo, Hyogo 679-5158, Japan}
\author{Takayuki Kiss}
\affiliation{Division of Materials Physics, Graduate School of Engineering Science, The University of Osaka, Toyonaka, Osaka 560-8531, Japan}
\author{Atsushi Higashiya}
\affiliation{RIKEN SPring-8 Center, Sayo, Hyogo 679-5158, Japan}
\affiliation{Faculty of Science and Engineering, Setsunan University, Neyagawa, Osaka 572-8508, Japan}
\author{Atsushi Yamasaki}
\affiliation{RIKEN SPring-8 Center, Sayo, Hyogo 679-5158, Japan}
\affiliation{Faculty of Science and Engineering, Konan University, Kobe, Hyogo 658-8501, Japan}
\author{Yuina Kanai-Nakata}
\author{Shin Imada}
\affiliation{RIKEN SPring-8 Center, Sayo, Hyogo 679-5158, Japan}
\affiliation{College of Science and Engineering, Ritsumeikan University, Kusatsu, Shiga 525-8577, Japan}
\author{Masaki Oura}
\author{Kenji Tamasaku}
\author{Makina Yabashi}
\author{Tetsuya Ishikawa}
\affiliation{RIKEN SPring-8 Center, Sayo, Hyogo 679-5158, Japan}
\author{Farid Labib}
\affiliation{Research Institute of Science and Technology, Tokyo University of Science, Tokyo 125-8585, Japan}
\author{Shintaro Suzuki }\thanks{Present address: Department of Physics, Aoyama Gakuin University, Sagamihara, Kanagawa 252-5258, Japan }
\author{Ryuji Tamura}
\affiliation{Department of Materials Science and Technology, Tokyo University of Science, Tokyo 125-8585, Japan}
\author{Akira Sekiyama} 
\affiliation{Division of Materials Physics, Graduate School of Engineering Science, The University of Osaka, Toyonaka, Osaka 560-8531, Japan}
\affiliation{RIKEN SPring-8 Center, Sayo, Hyogo 679-5158, Japan}
\affiliation{Spintronics Research Network Division, Institute for Open and Transdisciplinary Research Initiatives, The University of Osaka, Yamadaoka 2-1, Suita, Osaka, 565-0871, Japan}

\date{\today}

\begin{abstract}
We have investigated the orbital-dependent electronic states of $\rm{Cd_{6}}$Ce, a prototype of strongly correlated rare-earth-based Tsai-type quasicrystals and approximants (ACs) by soft and hard x-ray photoemission spectroscopy. Our results reveal that the 4$f$ orbitals are predominantly hybridized with the valence-band electrons far from the Fermi level ($E_F$), in sharp contrast to the hybridization with conduction electrons at $E_F$ seen for the intermetallic Ce-based compounds. This anomalous hybridization should be taken into account in discussing the unresolved magnetic ground state in $\rm{Cd_{6}}$Ce. These findings suggest that Cd-based ACs, some of which show the multi-step magnetic transitions, could provide a new platform for investigating novel magnetic properties that cannot be understood within the conventional framework of hybridization at $E_F$.

\end{abstract}
\maketitle

\section{Introduction}
Strongly correlated 4$f$-based electron systems exhibit fascinating quantum states like long-range magnetic ordering \cite{a,b}, unconventional superconductivity \cite{c,d}, and non-Fermi-liquid behavior \cite{e,f} in conventional periodic compounds. Understanding the mechanisms of these phenomena is one of the crucial issues in the 4$f$-based electron systems. Depending on the strength of hybridization between the localized 4$f$ orbitals and the itinerant valence bands ($c\mathchar`-f$ hybridization), the 4$f$ electrons are changed from the localized to itinerant character. Thus, one of the important keys to solving the issues is the hybridization effects.\par

$\rm{Cd_{6}}$M (M: Ca, Sr, Y, and rare-earth elements) 1/1 quasicrystal approximants (ACs) \cite{n,o,t,m,att} are periodic alloys composed of Tsai-type icosahedron clusters, which are regarded as the building blocks of the stable binary quasicrystals (QCs) $\rm{Cd_{5.7}}$Yb \cite{g} and $\rm{Cd_{5.7}}$Ca \cite{h}. Here, the clusters are periodically ordered for the ACs and quasiperiodically ordered for the QCs. Since the discovery of $\rm{Cd_{6}}$M, rare-earth-based QCs and ACs have been studied as the 4$f$-based strongly correlated electron systems. The first long-range magnetic ordering in ACs has been observed in $\rm{Cd_{6}}$Tb \cite{o}. After that, the antiferromagnetic transition has been reported in other $\rm{Cd_{6}}$R (R: Pr, Nd, Sm, Gd, Dy, Ho, Er, and Tm) \cite{m}. Moreover, in recent years, long-range magnetic ordering \cite{p,q}, non-Fermi-liquid behavior \cite{r}, and superconductivity \cite{s} have been discovered in the rare-earth-based ternary QCs and ACs, in which the ions in the Cd sites are replaced with two other elements. Therefore, $\rm{Cd_{6}}$M are regarded as prototypes for the rare-earth-based QCs and ACs. From this perspective, studying $\rm{Cd_{6}}$M is crucial for understanding the fundamental physical properties of the 4$f$-based electronic systems in QCs and ACs.\par

$\rm{Cd_{6}}$Ce, one of $\rm{Cd_{6}}$M, exhibits the orientational order of the tetrahedron at the center of the Tsai-type cluster below $\sim$ 460 K \cite{t}. It should be noted that the accurate stoichiometry composition ratio is $\rm{Cd_{37}}$$\rm{Ce_{6}}$, but it is referred to as $\rm{Cd_{6}}$Ce for simplicity \cite{n}. Macroscopically, $\rm{Cd_{6}}$Ce shows the magnetic transition around 0.5 K \cite{at}, but the details of this magnetic ground state of $\rm{Cd_{6}}$Ce are still unclear. Since the icosahedron formed of the Ce ions has a triangular structure, the geometrical frustration works on the Ce ions. Furthermore, the Ruderman-Kittel-Kasuya-Yosida (RKKY) interaction, which would be a dominant magnetic interaction, also acts between the Ce ions through the $c\mathchar`-f$ hybridization. To understand the magnetic ground state of $\rm{Cd_{6}}$Ce, the 4$f$ electronic states and the $c\mathchar`-f$ hybridization effects should be clarified. In the previous Ce 3$d$ core-level hard x-ray photoemission spectroscopy (HAXPES) on $\rm{Cd_{6}}$Ce \cite{u}, the localized 4$f$ electronic states have been revealed. However, the details about the $c\mathchar`-f$ hybridization effects in $\rm{Cd_{6}}$Ce are not clarified yet.  This motivates us to further investigate the valence electronic states, including the 4$f$ and other orbitals.\par

In this paper, we report on both Ce 4$f$ and other orbital valence electronic states of AC $\rm{Cd_{6}}$Ce. The Ce 3$d$-4$f$ resonance photoemission spectroscopy (RPES) directly probes the localized 4$f$ electronic states. Moreover, the $c\mathchar`-f$ hybridization strength at the Fermi level ($E_F$) is weaker than that in the crystalline Ce-based intermetallic Kondo systems.  The orbital contribution in the valence-band electronic states is discussed by using linearly polarized HAXPES, where it is revealed that the Cd 5$p$ orbital is dominant near $E_F$. Using the noncrossing approximation (NCA) calculation \cite{bbb,bbba,bbbc} based on the single-impurity Anderson model (SIAM) \cite{bbbd,bbbe}, it is found that the $c\mathchar`-f$ hybridization strength around 1 eV is much stronger than that at $E_F$, which differs from the $c\mathchar`-f$ hybridization effects in the intermetallic Ce-based compounds. \par

\section{Experimental details}
Single-crystalline AC $\rm{Cd_{6}}$Ce was prepared by the self-flux method \cite{at}. The Ce $M_5$-edge x-ray absorption spectroscopy (XAS) and Ce 3$d$-4$f$ RPES were carried out at BL17SU of SPring-8 \cite{vv}. The XAS spectrum was obtained in a total-electron-yield mode. The RPES spectra were measured by using a Scienta EW4000 electron analyzer. The overall energy resolution was set to 130 meV with the photon energy of $\sim$ 881 eV corresponding to the Ce $M_5$ edge. A circularly polarized (C-pol.) soft x-ray was used in both spectroscopies. The linearly polarized  HAXPES was performed at BL19LXU of SPring-8 with an MBS A1-HE hemispherical analyzer \cite{v}. The linear polarization of the incident hard x-ray with the photon energy of 7.9 keV was changed from the horizontal to the vertical direction by the double-diamond phase retarders. The degree of linear polarization
was estimated as $-0.95$, which corresponds to the linear polarization components along the horizontal and vertical directions of 2.5 and 97.5 \%, respectively. Since the hemispherical analyzer was positioned in the horizontal plane with an angle to incident photons of 60$\tcdegree$, the experimental configuration for the horizontally (vertically) polarized x-ray corresponds to the $p$-polarization ($s$-polarization). The overall energy resolution was set to 200 meV for the HAXPES measurement. The temperature was set to 20 K for XAS and RPES and 8 K for HAXPES. At these temperatures, AC $\rm{Cd_{6}}$Ce is in the paramagnetic phase. The clean surfaces were obtained by $\it in$-$\it situ$ fracturing for all measurements.\par

\section{RESULTS AND DISCUSSION}
\begin{figure}
	\begin{center}
	\includegraphics[keepaspectratio,,scale=0.52]{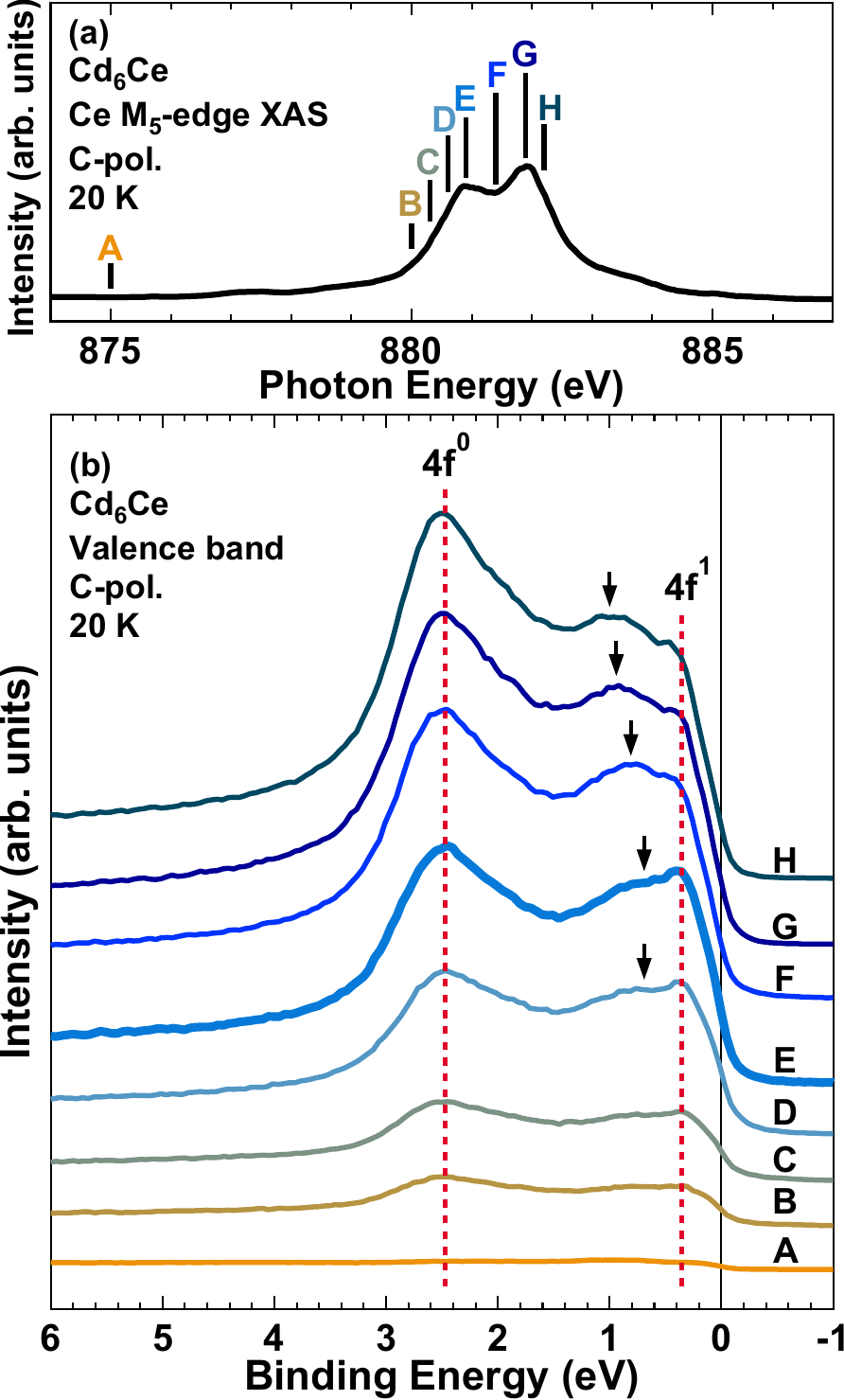}
	\caption{(a) Ce $M_5$-edge XAS spectrum of AC $\rm{Cd_{6}}$Ce. The labels A-H indicate the selected photon energies for the Ce 3$d$-4$f$ RPES of AC $\rm{Cd_{6}}$Ce. (b) Ce 3$d$-4$f$ RPES spectra of AC $\rm{Cd_{6}}$Ce at the photon energies A-H indicated in (a). The dashed lines indicate the peak binding energy of the $4f^0$ and $4f^1$ final states. The arrows show the hump structure of the $4f^1$ final states in the spectra at E-H.}
	\label{Fig1}
	\end{center}
\end{figure}

\begin{figure*}
	\begin{center}
	\includegraphics[keepaspectratio,,scale=0.62]{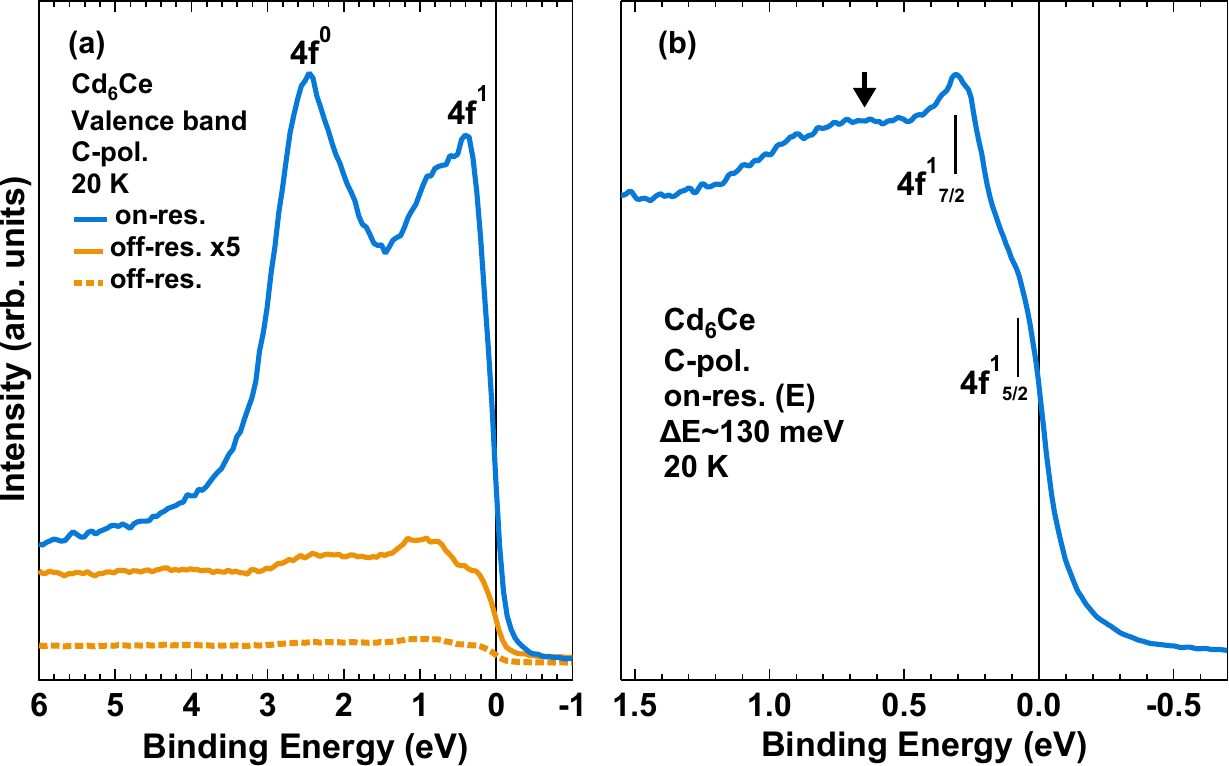}
	\caption{(a) On- and off-resonance (on-res., photon energy E and off-res., photon energy A indicated in Fig.~\ref{Fig1}(a)) valence-band photoemission spectra of AC $\rm{Cd_{6}}$Ce. (b) Enlarged view of the on-res.~spectrum of AC $\rm{Cd_{6}}$Ce with the energy resolution $\sim$ 130 meV. The solid lines indicate the $4f^1$$_{5/2}$ and $4f^1$$_{7/2}$ final states. The arrow shows the hump structure at around 0.7 eV.}
	\label{Fig2}
	\end{center}
\end{figure*}

\subsection{Ce $M_5$-edge XAS and 3$d$-4$f$ RPES}
Figure \ref{Fig1} shows the Ce $M_5$-edge XAS spectrum and the Ce 3$d$-4$f$ RPES spectra of AC $\rm{Cd_{6}}$Ce. The photon energies employed for the Ce 3$d$-4$f$ RPES are indicated by the vertical bold bars and labeled in Fig.~\ref{Fig1}(a). All valence-band spectra in Fig.~\ref{Fig1}(b) are normalized by the photon flux. Ongoing from the photon energy of A ($h\nu = 875$ eV) to E ($h\nu = 880.9$ eV), the evolution of two peaks at around 2.5 and 0.3 eV indicated by dashed lines is observed in the valence-band RPES spectra. The former is the $4f^0$ final state originating from the localized 4$f$ electronic state, whereas the latter is ascribed to the $4f^1$ final states possibly including the tail of the Kondo-resonance peak reflecting the $c\mathchar`-f$ hybridization effects \cite{w,aq}. As the photon energy changes from A to H, the peak binding energy of the $4f^0$ final state remains constant among the spectra at B-H. In addition, the peak binding energy of the $4f^1$ final states is also independent of the photon energy B to E. The constant binding energies of the $4f^0$ and $4f^1$ peaks indicate that the effect of possible Auger emission is negligible in the spectra. In the $4f^1$ spectral weight, the hump structure at around 0.7 eV is observed in the spectra at D and E, whereas this hump structure shifts toward the higher binding energy region in spectra at F-H. The kinetic energies of these hump structures also vary with the photon energy (see the Appendix). These results indicate that these structures in the spectra at D-H are mainly attributed to the resonance components. Hereafter, we refer to the spectrum at A as the off-resonance (off-res.) spectrum and that at E as the on-resonance (on-res.) spectrum.\par

Figure \ref{Fig2}(a) displays the on-res.~spectrum (at E) and the off-res.~spectrum (at A). The on-res.~spectrum shows the $4f^0$ and $4f^1$ peaks, whereas they are suppressed in the off-res.~spectrum. It is known that the relative intensity of the $4f^0$ final state to that of the $4f^1$ final states is related to the $c\mathchar`-f$ hybridization strength \cite{w,aq,ab,x}. Since the spectral weight of the $4f^0$ peak is larger than that of the $4f^1$ peak in Fig.~\ref{Fig2}(a), the $4f$ electrons in AC $\mathrm{Cd_{6}}$Ce have localized character. The bulk-sensitive Ce 3$d$ core-level HAXPES spectrum of AC $\mathrm{Cd_{6}}$Ce shows clear ionic-like on-site 3$d^9$4$f^1$ final-state multiplets, whereas 3$d^9$4$f^0$ peaks associated with itinerant 4$f$ states are negligible \cite{u}. Therefore, the result of the Ce 3$d$-4$f$ RPES is consistent with that of the Ce 3$d$ core-level HAXPES measurement.\par 

Figure \ref{Fig2}(b) shows the enlarged view of the on-res.~spectrum focused on the $4f^1$ final states to discuss the details of the $c\mathchar`-f$ hybridization effects in AC $\rm{Cd_{6}}$Ce. A shoulder structure at 0.08 eV and a peak structure at 0.3 eV are observed in Fig.~\ref{Fig2}(b). The former is the $4f^1$$_{5/2}$ final state corresponding to the so-called tail of the Kondo-resonance peak, whereas the latter is the $4f^1$$_{7/2}$ final state ascribed to its spin-orbit partner \cite{w,aq,ab}. As discussed in Fig. \ref{Fig1}(b), the hump structure is confirmed at around 0.7 eV. The details of this hump structure are discussed later. The spectral weight of the $4f^1$$_{5/2}$ final state is much suppressed compared with that of the $4f^1$$_{7/2}$ final state. In general, the $4f^1$$_{5/2}$ final states become more prominent compared with the $4f^1$$_{7/2}$ final states in the strong $c\mathchar`-f$ hybridization systems \cite{aq}. Therefore, the $c\mathchar`-f$ hybridization strength at $E_F$ is much weaker for AC $\rm{Cd_{6}}$Ce compared with that for the typical heavy-Fermion systems like Ce$\rm{Ru_{2}}$$\rm{Si_{2}}$ \cite{w}. \par

\begin{figure}
	\begin{center}
	\includegraphics[keepaspectratio,,scale=0.47]{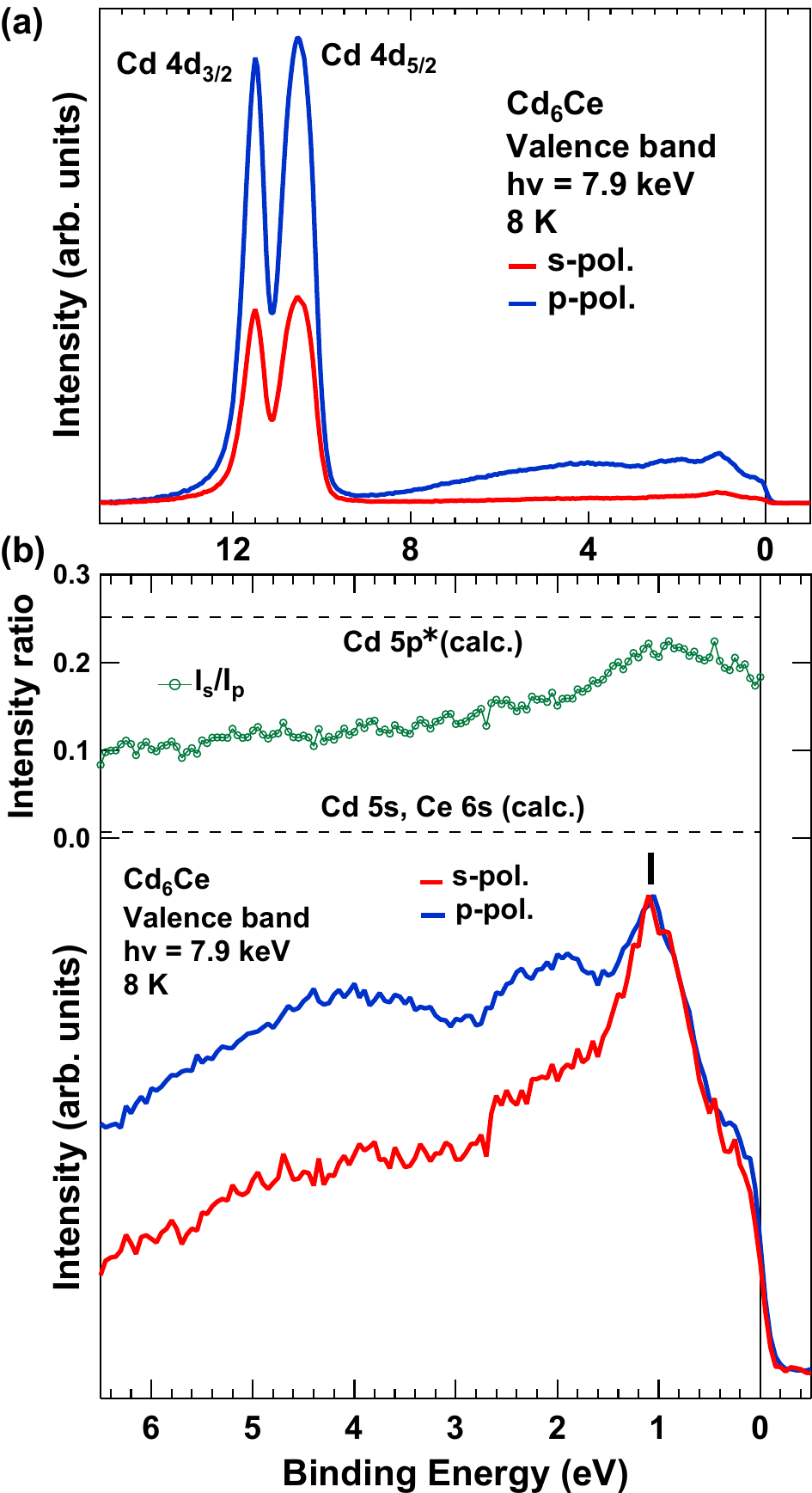}
	\caption{(a) Linearly polarized valence-band HAXPES spectra of AC $\rm{Cd_{6}}$Ce normalized by the photon flux. (b) Enlarged view of linearly polarized valence-band HAXPES spectra of AC $\rm{Cd_{6}}$Ce (bottom) and the photoelectron intensity ratio $I_s$/$I_p$ (top), which is calculated from the HAXPES spectra in (a). Both HAXPES spectra are normalized to the peak around 1 eV to enable a detailed comparison of the linear-polarization dependence of the spectral shape. The bold bar indicates the peak structure in both polarizations. The dashed lines in the upper panel show the calculated $I_s$/$I_p$ values listed in Table 1. An asterisk denotes the $I_s$/$I_p$ value of Cd 5$p$ orbital obtained with In 5$p$ calculation parameters, whereas unmarked values are calculated with the parameters for each orbital.}
	\label{Fig3}
	\end{center}
\end{figure}

\subsection{ Linearly polarized HAXPES}
While the Ce 3$d$-4$f$ RPES selectively probes the 4$f$ electronic states, the linearly polarized HAXPES \cite{ad} characterizes the orbital contributions to the valence band, which is essential for understanding the $c\mathchar`-f$ hybridization effects. Figure \ref{Fig3}(a) displays the linearly polarized valence-band HAXPES spectra of AC $\rm{Cd_{6}}$Ce. After the normalization by the photon flux, the Shirley-type backgrounds  \cite{ar} have been subtracted in both spectra. In both $p$- and $s$-polarization configurations, two peaks are observed at 11.5 and 10.6 eV, which are ascribed to the Cd $4d_{3/2}$ and $4d_{5/2}$ excitations, respectively. The spectral intensity of the $s$-polarization configuration ($I_s$) is suppressed compared with that of the $p$-polarization configuration ($I_p$) in Fig.~\ref{Fig3}(a). For the later discussion of the linear-polarization dependence of not only spectral intensity but also line shape, the lower panel of Fig.~\ref{Fig3}(b) shows the enlarged view of the valence-band HAXPES spectra for the $p$- and $s$-polarization configurations between 6.5 and $-0.5$ eV. Note that both spectra in Fig.~\ref{Fig3}(b) are normalized to the peak at around 1 eV. The hump structures centered at around 4 and 2 eV are observed in the $p$-polarization spectrum, which is suppressed in the $s$-polarization spectrum. Additionally, the spectral intensity in the $s$-polarization configuration is slightly weaker compared with that in the $p$-polarization configuration in the vicinity of $E_F$. \par

\begin{table}
\caption{Calculated photoelectron intensity ratio $I_{s}$/$I_{p}$ and photoionization cross sections $\sigma$ considering the composition ratio of AC $\rm{Cd_{6}}$Ce for Cd 4$d$, Cd 5$s$, Cd 5$p^*$, Ce 4$f$, Ce 6$s$, and Ce 5$d^*$ orbitals at the kinetic energy of 8 keV \cite{ag,ax}. Regarding the asterisks of Cd 5$p$ and Ce 5$d$ states, we have used the calculated $I_s$/$I_p$ for In 5$p$ and La 5$d$ orbitals because there are no calculation parameters for Cd 5$p$ and Ce 5$d$ orbitals.}
\label{Is_Ip}
\centering
\scalebox{1.0}{
 \begin{tabular}{ccccccc}\hline
 Orbital\hspace{2.2mm} &Cd 4$d$\hspace{2.2mm} &Cd 5$s$\hspace{2.2mm} & Cd 5$p^*$\hspace{2.2mm} &Ce 4$f$\hspace{2.2mm} &Ce 6$s$\hspace{2.2mm} &Ce 5$d^*$\\ \hline
$I_s$/$I_p$\hspace{2.2mm} &0.702\hspace{2.2mm} &0.007\hspace{2.2mm} &0.252\hspace{2.2mm} &0.989\hspace{2.2mm} &0.008\hspace{2.2mm} &0.546\\ \hline
$\sigma$ (b)\hspace{2.2mm} &47.6\hspace{2.2mm} &21.1\hspace{2.2mm} &9.5\hspace{2.2mm} &0.8\hspace{2.2mm} &1.4\hspace{2.2mm} &2.4\\ \hline
\end{tabular}}
\end{table}

\begin{figure*}
	\begin{center}
	\includegraphics[keepaspectratio,,scale=0.84]{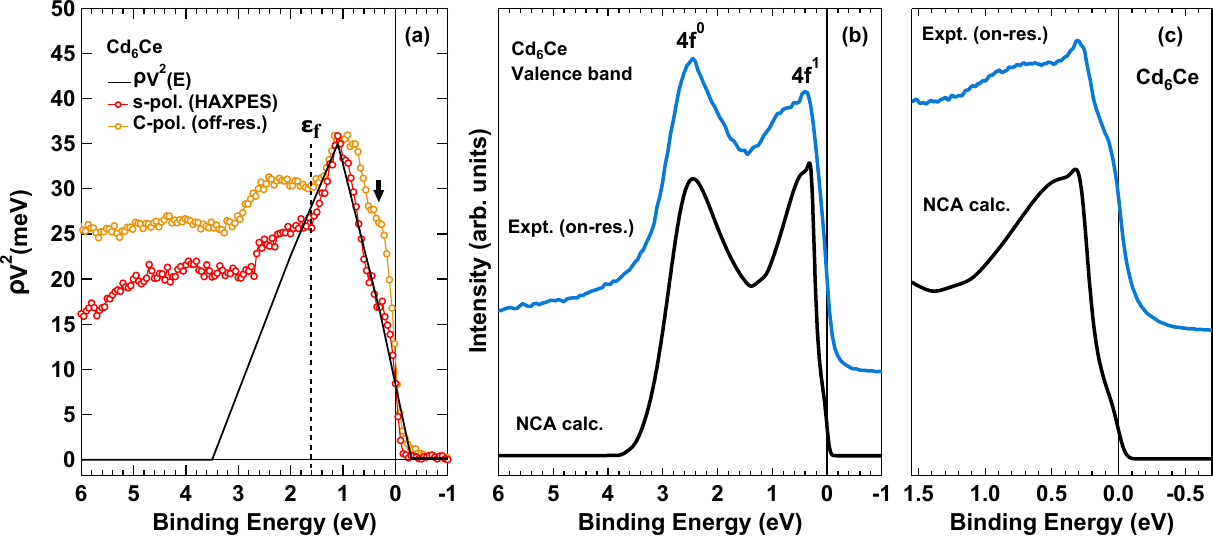}
	\caption{(a) Optimized $\rho$$V^2$$(E)$ in the NCA calculation compared with the off-res.~spectrum at  $h\nu = 875$ eV and valence-band HAXPES spectrum in the $s$-polarization configuration for AC $\rm{Cd_{6}}$Ce. The arrow shows the shoulder structure around 0.3 eV in the off-res.~spectrum. (b) Comparison of the on-res.~spectrum and that obtained by the NCA calculation. (c) Enlarged view of the on-res.~spectrum of AC $\rm{Cd_{6}}$Ce and the simulated spectrum from the NCA calculation.}
	\label{Fig4}
	\end{center}
\end{figure*}

To reveal the orbital components in the valence-band HAXPES spectra, in Fig.~\ref{Fig3}(b), we have compared the experimental photoelectron intensity ratio $I_s$/$I_p$, which has been obtained from the HAXPES spectra in Fig.~\ref{Fig3}(a) normalized by the photon flux, and the calculated ratio $I_s$/$I_p$ for the orbitals forming the valence bands listed in Table 1 \cite{ag,ax}. These calculated values are used to evaluate the dominant orbital contribution in each binding-energy region and to interpret the polarization dependence of the valence-band HAXPES spectra shown in the lower panel of Fig.~\ref{Fig3}(b). The experimental $I_s$/$I_p$ gradually increases from 6.5 to 2 eV and is estimated to range from $\sim$ 0.1 to $\sim$ 0.15. These values lie between the calculated $I_s$/$I_p$ values for the 5$p$ and $s$ orbitals. Thus, the contributions from Cd 5$p$ and $s$ orbitals are dominant in this region. On the other hand, the experimental $I_s$/$I_p$ rapidly increases from 2 eV with the peak at around 1 eV. The $I_s$/$I_p$ at 1 eV is experimentally estimated to be $\sim$ 0.21, which is close to the calculated $I_s$/$I_p$ value for the 5$p$ orbital. This result shows that the contribution of the Cd 5$p$ orbital is significant in this region, and the peak structure at around 1 eV observed in the valence-band HAXPES spectra for both polarization configurations is attributed to the Cd 5$p$ state. Additionally, the experimental $I_s$/$I_p$ gradually decreases toward $E_F$ and is evaluated to $\sim$ 0.18 at $E_F$, which is still larger than the calculated $I_s$/$I_p$ for the $s$ orbitals. It means that although the contribution of the Cd 5$p$ orbitals is reduced, it remains larger than that of the $s$ orbitals, even near $E_F$. It should be noted that, although we have used the parameters of In 5$p$ and La 5$d$ for the calculated $I_s$/$I_p$ for the Cd 5$p$ and Ce 5$d$ orbitals, respectively,  as an approximation, this conclusion is not sensitive to this approximation.\par

Densities of states (DOS) for $\rm{Cd_{6}}$M (M = Ca and Yb), calculated in the previous studies \cite{ah,ahh}, can be used as references for interpreting the results of linearly polarized HAXPES. The calculation shows that the Cd 5$p$ state is dominant near $E_F$ except for the narrow Yb 4$f$ band in $\rm{Cd_{6}}$Yb. Therefore, the significant contribution of the Cd 5$p$ orbital near $E_F$, revealed by the linearly polarized valence-band HAXPES, is in good agreement with the previous theoretical DOS of Cd-based ACs. It should be noted that the Ce 5$d$ contributions are minor in both spectra, which can be judged from the fact that the expected $I_s$/$I_p$ for the Ce 5$d$ orbitals is much larger than that for the Cd 5$p$ orbitals as shown in Table 1. \par

\subsection{NCA calculation}
For a more quantitative analysis of the $c\mathchar`-f$ hybridization effects in AC $\rm{Cd_{6}}$Ce, we have performed simulations by the NCA method \cite{bbb,bbba,bbbc} within SIAM \cite{bbbd,bbbe} using the orbital-contribution information from the linearly polarized HAXPES. The SIAM analysis is useful for evaluating the local $c\mathchar`-f$ hybridization effects, while the inter-site mutual 4$f$-4$f$ interactions are not directly taken into account. For the preparation of constructing the energy dependence of the hybridization strength $\rho$$V^2$$(E)$, we have compared the off-res.~spectrum at $h\nu = 875$ eV and the valence-band HAXPES spectrum in the $s$-polarization configuration, as shown in Fig.~\ref{Fig4}(a). The off-resonance spectrum shows the Cd 5$p$ peak structure around 1 eV, and a shoulder structure around 0.3 eV which is absent in the HAXPES spectra. The relative photoionization cross section of Ce 4$f$ orbital to Cd 5$p$ orbital ($\rm{\sigma_{Ce}}$$\,$$_{4f}$/$\rm{\sigma_{Cd}}$$\,$$_{5p}$) is estimated to be $\sim$ 6 in the off-res.~spectrum and $\sim$ 0.08 in the HAXPES spectrum \cite{ax,axx}. In addition, the binding energy of this shoulder structure is very close to that of the $4f^1$$_{7/2}$ final state in the on-res.~spectrum. These results identify the shoulder around 0.3 eV in the off-res.~spectrum as the Ce 4$f$ state.\par

Based on the relatively large contribution of the Cd 5$p$ orbital, we assume that the 4$f$ states primarily hybridize with the Cd 5$p$ orbital. Thus, the $\rho$$V^2$$(E)$ shown in Fig.~\ref{Fig4}(a) is optimized to be roughly proportional to the valence-band HAXPES spectrum in the $s$-polarization configuration, where the Cd 5$p$ orbital contribution is dominant while that of other orbitals is suppressed. This $\rho$$V^2$$(E)$ possesses the maximum hybridization strength at 1.1 eV and linearly decreases toward $E_F$. The value of $\rho$$V^2$$(E)$ is 35 meV at 1.1 eV and  8.3 meV at $E_F$. The bare 4$f$ binding energy ($\epsilon_f$) is set to 1.61 eV. For the simulations of the 4$f$ spectrum, the experimental energy resolution is taken into account.\color{black}\par

Figure \ref{Fig4}(b) shows the comparison of the on-res.~spectrum and the simulated 4$f$ spectrum, in which the intensity of the $4f^0$ peak is comparable to that of the $4f^1$ peak. As shown in the figure, we have semiquantitatively reproduced the experimental Ce 3$d$-4$f$ RPES spectra by the NCA calculation using the parameters listed above. Figure \ref{Fig4}(c) displays the enlarged view of the comparison of the experimental on-res.~spectrum with the NCA-simulated spectrum. The intensity of the $4f^1$$_{5/2}$ final state is suppressed compared with that of the $4f^1$$_{7/2}$ final state also in the simulation. Moreover, the hump structure at around the binding energy of 0.7 eV is well reproduced in the spectrum of the NCA calculation. \par

Reproducing the on-res.~$4f$ spectral weight from $E_F$ to 1 eV requires the energy-dependent $c$–$f$ hybridization strength. Specifically, it is suppressed at $E_F$ but strong near 1.1 eV. Therefore, it is revealed that the 4$f$ orbitals in AC $\rm{Cd_{6}}$Ce are primarily hybridized with the Cd 5$p$ orbital centered at around 1 eV, rather than with the conduction bands located at $E_F$, which leads to the suppressed $4f^1$$_{5/2}$ final state and the hump structure at around 0.7 eV. Here, the weak $4f^1$$_{5/2}$ final state and the hump structure at the higher binding energy side of the $4f^1$$_{7/2}$ component have also been observed in the Ce 3$d$-4$f$ RPES spectra of the Ce pnictide compounds \cite{ab}, which have been known as low-carrier systems with very low density of states at $E_F$.  From the NCA calculation for these Ce pnictide compounds, the $c\mathchar`-f$ hybridization strength around 1-2 eV is much larger than that around $E_F$ \cite{ab}. This $c\mathchar`-f$ hybridization effect in AC $\rm{Cd_{6}}$Ce is substantially different from that observed in the intermetallic Ce-based compounds showing the Kondo effect, where the 4$f$ orbitals are strongly hybridized with the conduction electrons at $E_F$ \cite{aq}. It should be noted that the $4f^1$ final states in AC $\rm{Cd_{6}}$Ce can take several different states because the 4$f$ orbitals are hybridized with the energetically continuous valence bands. The resonance enhancements of these states can depend on photon energy. This would explain the shift of the hump structures in spectra at F-H of Fig.~\ref{Fig1}(b).

\subsection{Discussion}
From the comparison of the experimental on-res.~spectra and the calculated spectrum, it is revealed that the 4$f$ orbitals are mainly hybridized with the Cd 5$p$ band located at around 1 eV, rather than with the conduction bands at $E_F$. This $c\mathchar`-f$ hybridization effect is unusual compared with that in the Ce-based intermetallic compounds in which the hybridization at $E_F$ is dominant. In terms of orbital character, although the experimental verification of the detailed energy-dependent orbital symmetry of the valence bands from $E_F$ to 1 eV is difficult,  the orbital symmetry of the valence bands at 1 eV would be more suitable for the $c\mathchar`-f$ hybridization than that of the bands at $E_F$. It would be responsible for this unique $c\mathchar`-f$ hybridization effect in AC $\rm{Cd_{6}}$Ce.\par

Because the $c\mathchar`-f$ hybridization strength at $E_F$ is suppressed in AC $\rm{Cd_{6}}$Ce, the screening of the Ce 4$f$ magnetic moments by the conduction electrons hardly occurs, and thus the 4$f$ electrons have the magnetic moments. The RKKY interaction between these magnetic 4$f$ electrons is likely to lead to the magnetic transition at around 0.5 K in  AC $\rm{Cd_{6}}$Ce. As an analogy, a closely related situation to that observed for AC $\rm{Cd_{6}}$Ce has been discussed for CeSb, which shows the complex multiple antiferromagnetic transitions with various kinds of magnetic structures \cite{aag}. In the Ce 4$d$-4$f$ RPES on CeSb, the suppression of the $4f^1$$_{5/2}$ final states and the appearance of a peak structure on the higher binding energy side of the $4f^1$$_{7/2}$ component have been observed \cite{aae}. In addition, the theoretical calculation shows that the $c\mathchar`-f$ hybridization strength at $E_F$ is weaker than that at the higher binding energy side in CeSb \cite{aaf}. This similarity to CeSb suggests that the observed anomalous $c\mathchar`-f$ hybridization is important for understanding the magnetic ground state of AC $\rm{Cd_{6}}$Ce. It should be noted that the magnetic ground state should also be affected by the inter-site 4$f$-4$f$ interaction and the geometric frustration, which are not taken into account in this study.\par

It has been reported that some of $\rm{Cd_{6}}$M exhibit multi-step magnetic transitions \cite{o,m}. However, the mechanism of these transitions has still remained unclear. Our study introduces an additional viewpoint based on the unique $c\mathchar`-f$ hybridization effect, which has not been considered in the previous discussions of the magnetism of the $\rm{Cd_{6}}$M system. By analogy with CeSb, we speculate that, if similar anomalous $c\mathchar`-f$ hybridization is also realized in other $\rm{Cd_{6}}$M, it might be relevant to their complex magnetic behavior. In this sense, $\rm{Cd_{6}}$M could offer a new platform for exploring exotic magnetic properties beyond the conventional framework of  $c\mathchar`-f$ hybridization effect at $E_F$, which has been typically seen in many rare-earth intermetallic compounds.

\section{Summary}
In summary, we have performed the Ce 3$d$-4$f$ RPES and the linearly polarized HAXPES on AC $\rm{Cd_{6}}$Ce. The 4$f$ electrons are localized, and the $c\mathchar`-f$ hybridization strength at $E_F$ is much weaker compared with the Kondo systems. In addition, the Cd 5$p$ orbital has a large contribution peaking at around 1 eV in the valence-band HAXPES spectra, which is consistent with the previous theoretical calculations of Cd-based ACs. From the comparison of the on-res.~spectrum of the Ce 3$d$-4$f$ RPES and the calculated spectrum, the 4$f$ orbitals mainly hybridize with the Cd 5$p$ bands centered at around the binding energy of 1 eV, rather than with the conduction bands at $E_F$. This unique $c\mathchar`-f$ hybridization effect could provide a new perspective on the magnetic properties of the Cd-based ACs.\par

\section{ACKNOWLEDGMENTS}
We acknowledge H. Hashizume, Y. Shimada, T. Usui, T. Miyazaki, Y. Chen, Y. Murakami, and N. Tanaka for supporting the HAXPES measurement. We also thank A. Enomoto and Y. Torii for the support of the Ce 3$d$-4$f$ RPES measurement. The measurements at BL19LXU (Proposals Nos. 20210068 and 20220071) and those at BL17SU (Proposals Nos. 20210082 and 20220027) were performed under the approval of RIKEN at SPring-8. This work was financially supported by a Grant-in-Aid for Innovative Areas (JP19H05817, JP19H05818, JP20H05271, and JP22H04594), a Grant-in-Aid for Transformative Research (JP23H04867), a Grant-in-Aid for Scientific Research (JP19K14663, JP20K20900, JP22K03527, and JP24K03202) from JSPS and MEXT, and CREST (JPMJCR22O3) from JST. G. Nozue was supported by the University of Osaka fellowship program of Super Hierarchical Materials Science Program and by the JSPS Research Fellowships for Young Scientists. \par

\section{APPENDIX: Ce 3$d$-4$f$ RPES spectra with kinetic energy scale}
\begin{figure}
	\begin{center}
	\includegraphics[keepaspectratio,,scale=0.55]{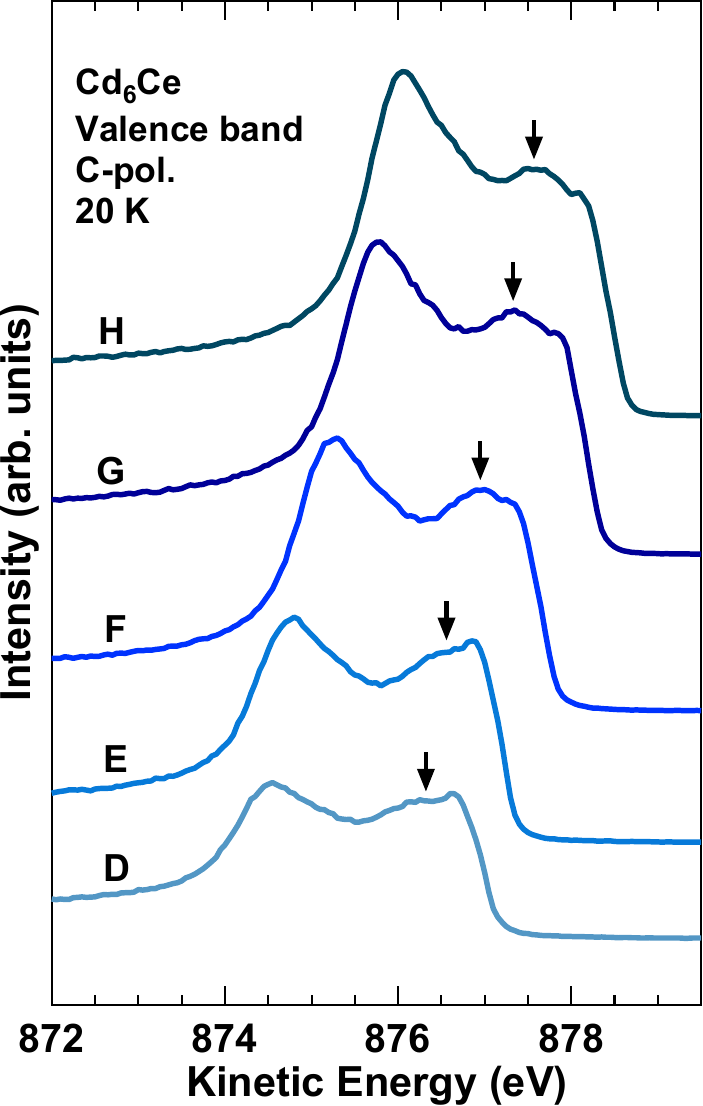}
	\caption{Ce 3$d$-4$f$ RPES spectra of AC $\rm{Cd_{6}}$Ce at the photon energies D-H indicated in Figure \ref{Fig1}(a) as a function of the photoelectron kinetic energy. The arrows show the hump structure of the $4f^1$ final states in the spectra at D-H.}
	\label{Fig5}
	\end{center}
\end{figure}

Figure \ref{Fig5} shows the Ce 3$d$-4$f$ RPES spectra of AC $\rm{Cd_{6}}$Ce as a function of the photoelectron kinetic energy at the photon energies D-H labeled in Fig.~\ref{Fig1}(a). The kinetic energy of the hump structures in the spectra at D-H is shifted toward the higher kinetic energy region by changing the photon energy in Fig.~\ref{Fig5}. In the case of the Auger emission, the kinetic energy of the structure should be constant regardless of photon energy. Therefore, these results indicate that these structures in the spectra at D-H are mainly contributed by the resonance photoemission components.

\end{document}